# High-rate Plastic Deformation of Nanocrystalline Tantalum to Large Strains: Molecular Dynamics Simulation

Robert E. Rudd

Lawrence Livermore National Laboratory, 7000 East Ave., L-045, Livermore, CA 94550 USA

February 13, 2009

**Abstract.** Recent advances in the ability to generate extremes of pressure and temperature in dynamic experiments and to probe the response of materials has motivated the need for special materials optimized for those conditions as well as a need for a much deeper understanding of the behavior of materials subjected to high pressure and/or temperature. Of particular importance is the understanding of rate effects at the extremely high rates encountered in those experiments, especially with the next generation of laser drives such as at the National Ignition Facility. Here we use large-scale molecular dynamics (MD) simulations of the high-rate deformation of nanocrystalline tantalum to investigate the processes associated with plastic deformation for strains up to 100%. We use initial atomic configurations that were produced through simulations of solidification in the work of Streitz et al [Phys. Rev. Lett. 96, (2006) 225701]. These 3D polycrystalline systems have typical grain sizes of 10-20 nm. We also study a rapidly quenched liquid (amorphous solid) tantalum. We apply a constant volume (isochoric), constant temperature (isothermal) shear deformation over a range of strain rates, and compute the resulting stress-strain curves to large strains for both uniaxial and biaxial compression. We study the rate dependence and identify plastic deformation mechanisms. The identification of the mechanisms is facilitated through a novel technique that computes the local grain orientation, returning it as a quaternion for each atom. This analysis technique is robust and fast, and has been used to compute the orientations on the fly during our parallel MD simulations on supercomputers. We find both dislocation and twinning processes are important, and they interact in the weak strain hardening in these extremely fine-grained microstructures.

## The Plasticity of Nanocrystalline Metals at High Rates

The mechanical behavior of nanocrystalline materials has been the subject of numerous investigations. The research has been motivated in part by the increased strength of fine-grained materials due to the Hall-Petch effect [1] and facilitated by the great strides made in producing full density materials with extremely fine-grained microstructures. Surprising effects have been found, such as the inverse Hall-Petch effect in which the strength decreases with decreasing grain size for sufficiently small (nanoscale) grains [2,3,4,5] and unconventional dislocation processes due to the interaction of partials with the plethora of grain boundaries [6,7]. In addition to the conventional and unconventional dislocation processes, nanocrystalline materials are thought to undergo grain boundary sliding and grain rotation if the grains are sufficiently small [3]. While the nanocrystalline materials can be very strong, their low ductility and concomitant low fracture toughness has been an issue [8]. For a recent review of the mechanical behavior of nanocrystalline materials see Ref. [9], and for a review emphasizing modeling see Ref. [10].

Most of the envisaged applications of nanocrystalline metals have been in engineering applications with low rates, but there is growing interest in using them as structural materials for applications in which they are, or may be, subjected to very high rate deformation. For example, some capsule designs for inertial confinement fusion need high strength materials to contain the deuterium-tritium (DT) fuel as it is loaded, and then need an understanding of the strength during the laser-driven compression. The thin shell material also needs to be very isotropic elastically in order for the compression waves to retain the symmetry of the spherical capsule, avoiding the Rayleigh-Taylor instabilities that can spoil ignition. Ultra-fine-grained microstructures are one way to attain the needed strength and isotropy. Nanocrystalline gold-copper [11] and nanocrystalline tantalum [12,13] have been investigated for high power laser applications.

The understanding of high-rate plastic deformation, such as that induced by strong laser [14,15,16,17] and Z-pinch [18,19] drives, poses considerable challenges. The conditions are far from the ambient conditions of most mechanical tests. The plastic deformation rates in these experiments can be at least as high as $10^8$/s [15], and the materials are often subject to high pressure up to megabars [20] and large plastic strains as well. Shock wave experiments also typically induce a large temperature rise due to the entropy generation at the shock front as dictated by the Rankine-Hugoniot equations; more recent ramp-wave techniques have provided a means of achieving high pressure and high rates with substantially less heating, keeping the systems solid up to higher pressures [16,18]. Only a limited number of probes is available to determine the how the material behaves under these dynamic loads. Velocity Interferometer System for Any Reflector (VISAR) measurements determine the particle velocity [21]. Stepped samples can be used with VISAR to determine wave transit times. These VISAR and related techniques can be used to infer the stress-density relations of the material [16]. X-ray diffraction can provide insight into dislocation density [22], phase transitions [23] and plastic relaxation [24,25]. Optical pyrometry and a few other diagnostics can be used to determine the temperature under the right conditions. Some experiments have investigated the strength of conventional polycrystalline Ta as a function of loading rate, but the characterization of the plasticity in situ has been very limited [26,27,28]. The available probes provide an incomplete picture of the response of materials to high rate loading, so the ability to complement these experimental data with computer models is very important.

Several approaches have been taken to model the plasticity associated with high-rate deformation of metals. Typically either conventional polycrystalline materials or single crystals have been studied. Several continuum-level models have been developed for high-rate material strength including Ta [29,30,31,32] and plastic relaxation processes in high-rate deformation [33,34]. More recently, MD has been used to study the plasticity in ramp and shock waves [35,36] and the generation of nanocrystalline systems following a phase transition at a shock front [37]. One MD study considered the strength of nanocrystalline copper under shock loading [38].

Here we use MD simulations to investigate the rate dependence of plastic deformation in nanocrystalline Ta at the high rates associated with laser and Z-pinch drives. Because these systems have a ramp or quasi-isentropic loading, we focus on the response of a representative volume element rather than modeling the wave front explicitly. The wave front in these systems is spread over a thickness $L_{wave} = c\varepsilon_{max} / \dot{\varepsilon}$ where $c$ is the wave velocity, $\varepsilon_{max}$ is the ultimate strain, and $\dot{\varepsilon}$ is the strain rate. With $c$=3400 m/s, $\varepsilon_{max}$=0.1, and $\dot{\varepsilon}$=$10^8$/s, the wave front thickness is 3.4 microns, very large for an MD simulation; so instead of modeling the wave explicitly we consider how a nanocrystalline MD system responds to a uniform deformation at specified rates. We analyze the rate dependence of the stress-strain response and consider the mechanisms of the plastic deformation.

**Simulation Technique**

We have used classical MD based on the Finnis-Sinclair potential for tantalum [39,40]. This particular potential was chosen to provide a good representation of the elastic properties of Ta at a moderate computational cost. Another potential, the Model Generalized Pseudopotential Theory (MGPT) potential, has been shown to be more accurate for high pressure applications [41], but the Finnis-Sinclair potential is widely used, much less expensive than MGPT and is suitable for our purposes here. The interatomic forces on each atom were determined from minus the gradient of the potential, and Newton's equations of motion ($F=ma$) were integrated in time using the velocity Verlet integrator [42] with a time step of 3 fs. The work done on the system by the shear deformation was compensated by a thermostat, implemented through velocity renormalization. The required cooling power is quite large. The plastic work rate at a strain rate of $10^9$/s against a flow stress of 6 GPa is 0.7 eV/atom/ns, for example, and this heat must be removed by the thermostat in the isothermal simulation. In dynamic experiments the plastic work rate can cause appreciable heating in shock-free ramp loading conditions which, free of shocks, might be expected to be quasi-isentropic [20]. Our code FEMD, a concurrent molecular dynamics/finite element modeling code [43], was run on a supercomputer in pure MD mode, using 128 processors in an 8x4x4 spatial domain decomposition for the higher strain rates, and 512 processors in an 8x8x8 spatial domain decomposition for the lower strain rates on the MCR, Zeus and Atlas supercomputers [44]. Quantities such as the stress tensor, the temperature, the total energy and so forth were computed and stored throughout the simulation. Also, the grain orientations were computed on the fly every 100 time steps and written to files for visualization and further analysis.

For the initial configuration of the atoms in the simulation, we use a nanocrystalline Ta microstructure that was produced in prior work by Streitz et al. through the rapid compression of molten tantalum [45,46]. In their MD simulations a 16384000 atom liquid tantalum system was rapidly compressed from the liquid region of the phase diagram to the body-centered cubic (bcc) solid region of the phase diagram. According to the conventional Becker-Döring nucleation theory, this solidification process takes place through nucleation of crystallites that are sufficiently large to overcome the cost of the interfacial energy, and these critical nuclei grow [47]. The simulations had neither surfaces nor other distinct heterogeneities, so the crystallite growth began with a homogeneous nucleation process from the bulk fluid. It has been argued that the length scale of the microstructure should decrease with increasing compression rate according to $L \propto \dot{\varepsilon}^{-1/4}$ [48,49]: the high compression rate in these simulations led to a nanocrystalline microstructure with the larger grains 10-20 nm across and many smaller grains. The microstructure did not have a significant texture, and the departures from perfectly random orientation can be attributed to the statistics resulting from a fairly small number (a few tens) of larger grains. Dynamic compression experiments may also compress fluids rapidly enough to produce a nanocrystalline solid, just as the simulations of Streitz et al. [45] did, although the corresponding experiments on Ta have yet to be done.

We choose to study the system produced through this pressure-induced solidification process partly because it should be a more realistic representation of what may be encountered in dynamic experiments, and partly to compare and contrast with simulations based on microstructures that are constructed, for example using a Voronoi technique [50], although such a comparison is beyond the scope of this article. The initial microstructure has a characteristic size of ~10-20 nm with a grain orientation distribution that is roughly random (the pole plot is not completely uniform, but we attribute the weak non-uniformity to the poor statistics of relative large grains). The grains are largely free of dislocations and other defects, apart from the grain boundaries. The initial grain boundaries include some twin boundaries, but most are random boundaries [51].

The initial microstructure has then been subjected to a volume-conserving (isochoric) shear strain by changing the shape of the periodic simulation box at a specified strain rate. We consider both uniaxial compression and biaxial compression. They are complementary due to the constant volume: uniaxial compression is necessarily biaxial dilation, and vice-versa. Both involve tetragonal deformations of the simulation box. Uniaxial compressive strain is common in dynamic experiments such as plate impact experiments [21] and the more recent ramp-wave experiments [16,18]. Typically, the strain is purely uniaxial with the transverse directions unstrained, rather than the isochoric uniaxial strain studied here. We choose an isochoric deformation to focus on the shear strain behavior separate from pressurization effects. Biaxial compressive strain is also encountered in some dynamic experiments, such as in the compression of laser-driven inertial confinement fusion capsules planned for the National Ignition Facility [52].

The deformation of the system was imposed by a slight, instantaneous change in the box shape at each time step, imposed through scaling the metric while keeping the scaled coordinates of the atoms unchanged. Specifically, the atomic coordinates for atom $\mu$, $x_{\mu i}$, are related to the scaled coordinates, $s_{\mu j}$, running from 0 to 1 by the metric $h_{ij}$. This approach maintains the periodic boundary conditions in all three dimensions. The technique is akin to the scaling used in the constant-pressure technique of Parrinello and Rahman [53]. The form of the metric is a diagonal matrix with $\{1/\sqrt{h_{33}(t)}, 1/\sqrt{h_{33}(t)}, h_{33}(t)\}$ on the diagonal so that the volume is constant. The metric $h_{33}(t)$ evolves at a constant engineering strain rate, so $h_{33}(t) = h_{33}(t=0) (1 + \dot{\varepsilon} t)$ where $\dot{\varepsilon}$ is the strain rate. Here $h_{33}(t=0)$ is the length of a side of the initial cubic box size: 66.8 nm. We verified the plastic isotropy of the nanocrystalline system by running simulations in which the distinguished axis is in other directions (e.g. compressing in $x$ rather than $z$). Only the $z$-oriented simulations are reported here.

**Analysis of the Microstructure**

The plastic deformation is expected to be affected by, and to have an effect on, the grain structure of the nanocrystalline system. In order to observe these interactions we need a technique to monitor the microstructure, and to quantify its changes in a computationally efficient way. Here we take an approach that assigns to each atom an orientation that describes the atom's local environment. Like any rigid body rotation, the lattice orientation is specified by three numbers, often expressed in terms of Euler angles [54]. Here we find it convenient to use a quaternion to describe the orientation. A quaternion may be expressed as a 4-vector ($q_0, q_1, q_2, q_3$) with the constraint that sum $\sum q_i^2 = 1$ for unit quaternions. They obey multiplication rules that are appropriate for rotations, through their relation to the spinor representation of the rotation group SO(3). In rigid body mechanics quaternions have been used to provide a representation of the orientation of the rigid body that is free from coordinate singularities [55,56,42], and they are known to provide the basis for an alternative formulation of quantum mechanics [57]. They provide a powerful way to express the orientations and misorientations of a polycrystalline material, facilitating the analysis of the grain boundary network including coincident-site lattice properties [55,58], and it is for this reason we use them here. Based on the knowledge of any two directions in the lattice, e.g. $[111]$ and $[\bar{1}11]$, the lattice orientation and the corresponding unit quaternion can be determined.

Here we calculate the quaternion in three steps: (1) find $\hat{n}$ and $\hat{m}$, unit 3-vectors at each atom that point along particular directions in the lattice, (2) determine the corresponding Euler angles $(\theta, \phi, \psi)$, and (3) map these onto the quaternion in the usual way [56]:

$$q_0 = \cos(\theta/2) \cos[(\varphi + \psi)/2], \quad q_1 = \sin(\theta/2) \cos[(\varphi - \psi)/2],$$

$$q_2 = \sin(\theta/2) \sin[(\varphi - \psi)/2], \quad q_3 = \cos(\theta/2) \sin[(\varphi + \psi)/2]. \quad (1)$$

In particular we take $\hat{n}$ and $\hat{m}$ to point along <111> axes of the local bcc crystal lattice of Ta in these simulations for perfect lattice sites. Effectively $\hat{n}$ determines the first two Euler angles $(\theta, \phi)$ and the position of $\hat{m}$ with respect to $\hat{n}$ determines $\psi$. The algorithm to find the quaternion is presented below based on the configuration of the nearest neighbor atoms: neighbors that form a cube in the perfect bcc lattice with corners in the $\langle 111 \rangle$ directions. The algorithm returns a well-defined quaternion for any atom, even if the local lattice is not a perfect crystal such as at grain boundaries and dislocation cores.

The algorithm begins by finding a set of 8 near neighbor atoms. The number 8 is chosen to equal the number of nearest neighbors in the strain-free perfect bcc lattice, and this number would be different in different lattices. Consider the set of 8 unit vectors $\hat{n}_i$ pointing from the center in the direction to the 8 nearest atoms. In a perfect lattice these are the <111> directions. Here the center is taken to be the center of mass of the 8 neighbors rather than the central atom location because it reduces the thermal noise, a technique we introduced to calculate centrosymmetry deviation for dislocation detection [41]. The vector $\hat{n}$ is taken to be the unit vector $\hat{n}_i$ most in the $[111]_{box}$ direction; i.e. the vector $\hat{n}_i$ of the eight such that $\hat{n}_{ix} + \hat{n}_{iy} + \hat{n}_{iz}$ is maximal. The second vector $\hat{m}$ is taken to be the one of the remaining seven $\hat{n}_i$ most in the $[\bar{1}11]_{box}$ direction; i.e., the one such that $-\hat{n}_{ix} + \hat{n}_{iy} + \hat{n}_{iz}$ is maximal. The subscript "box" indicates that these directions are with respect to the simulation box, rather than the local lattice (which at this point is not yet determined), and the notation $\hat{e}_{111}^{box}$ will be used to denote the corresponding unit vector. The directions $[111]_{box}$ and $[\bar{1}11]_{box}$ are somewhat arbitrary, but picking some definite directions helps to ensure that the orientation is determined consistently throughout the system. Even at lattice sites free of any defect, at finite temperature $\hat{n}$ and $\hat{m}$ do not point exactly along the lattice $\langle 111 \rangle$ directions due to thermal fluctuations. Averaging over the neighbors can help to reduce the noise.

We take $\hat{n}$ to define the lattice $[111]$ direction: $\hat{e}_{111} = \hat{n}$. We then define $\hat{e}_{0\bar{1}1} = \hat{n} \times \hat{m} / |\hat{n} \times \hat{m}|$. From these two lattice vectors all of the other directions can be determined: e.g. $\hat{e}_{2\bar{1}\bar{1}} = \hat{e}_{111} \times \hat{e}_{0\bar{1}1}$ and $\hat{e}_{001} = 3^{-1/2} \hat{e}_{111} + 2^{-1/2} \hat{e}_{0\bar{1}1} - 6^{-1/2} \hat{e}_{2\bar{1}\bar{1}}$. The first two Euler angles are $\theta = \arccos(\hat{e}_{111} \cdot \hat{e}_{111}^{box})$ and $\phi = \arcsin(\hat{e}_{111} \cdot \hat{e}_{0\bar{1}1}^{box} / \sin\theta)$. The third Euler angle is then given by $\psi = \arcsin(\hat{e}_{0\bar{1}1} \cdot \hat{e}_{111}^{box} / \sin\theta)$. On the multivalued meridian $\theta = 0$, we take $\phi = 0$ and $\psi = \arccos(\hat{e}_{0\bar{1}1} \cdot \hat{e}_{0\bar{1}1}^{box})$. Due to the cubic symmetry of the lattice, the Euler angles have restricted ranges: e.g., the $\psi$ range is limited because of the three-fold axis of the cubic crystal [59]. The quaternion is then calculated from Eq. (1).

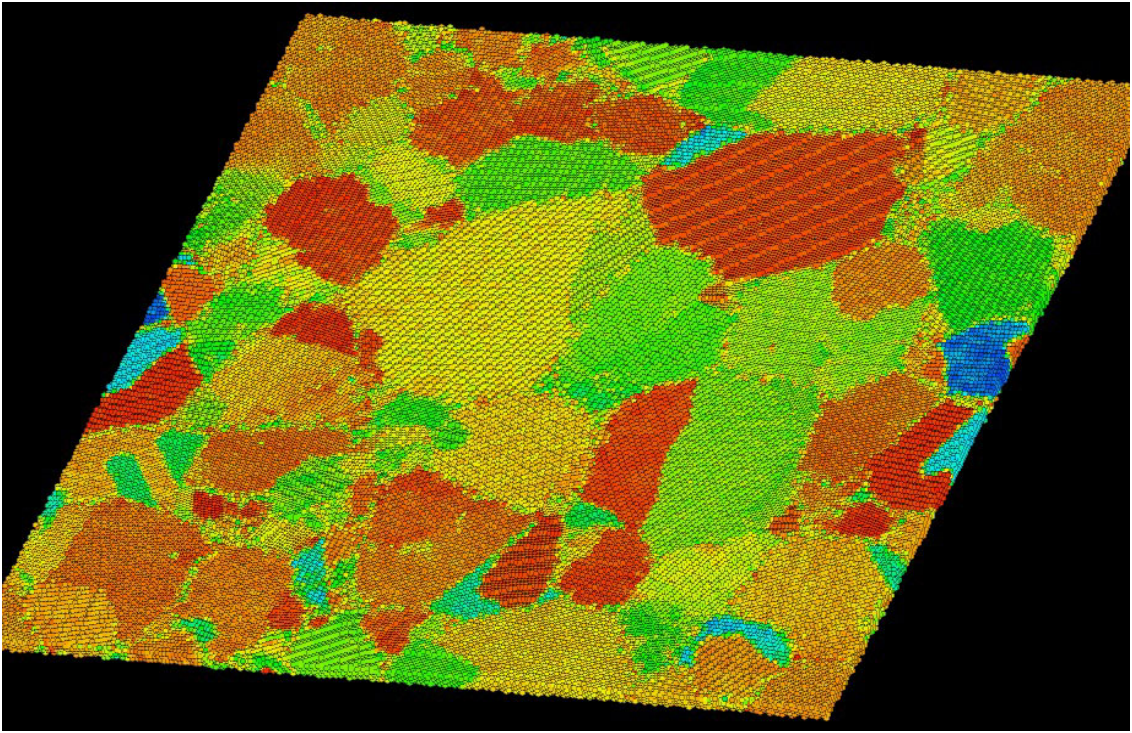

Figure 1. Nanocrystalline Ta microstructure at ε=0.02. A slice of the simulation box is shown with the atoms colored according to the quaternion orientation as described in the text. Initially the box is 66.8 nm across, and the larger grains are 10-20 nm.

This algorithm has been implemented to run on parallel processors with distributed memory. It makes use of the ghost atoms (from neighboring processors) and neighbor lists already generated for the interatomic force calculations. For visualization purposes, we follow one of the practices in orientation imaging microscopy (OIM) [60] such as to map the spatial part of the quaternion vector ($q_1,q_2,q_3$) to the RGB color such that each of the three components is mapped to the intensity of a color: red, green or blue. We also have used the orientation colormap sending $(3^{-1/2}\sin\psi + 1/2)|\hat{n}|$ to RGB (where $|\hat{n}|$ denotes the absolute value of the components of $\hat{n}$, not its norm), which is the technique we have used for the figures in this article. In practice this approach works well for visualization, even though it does not use all of the dynamic range of the RGB color spectrum. It can lead to some boundary artifacts since the fundamental domain of the quaternions for a cubic lattice is only a small subset of the possible range of quaternions, but artifacts have not been a significant problem in our analysis in the simulations run to date. A visualization of the microstructure early in the biaxial simulation using this orientation analysis is shown in Fig. 1.

**Simulation Results**

**Biaxial Compression**. First we consider the case of biaxial compression. The system begins in a state of zero shear stress and proceeds through several stages of deformation during the compression: elastic shear, the first departures from elasticity as plastic deformation commences, an over-driven regime in which the flow stress rises above its ultimate plateau, a peak in the flow stress, flow stress relaxation, and finally a weak strain hardening regime. The stress-strain curve is plotted in Fig. 2, showing these different regimes. The shear strain, here and in what follows, is calculated according to the formula:

$$\varepsilon = \frac{\sqrt{2}}{3}\left[(\varepsilon_{xx}-\varepsilon_{yy})^2 + (\varepsilon_{yy}-\varepsilon_{zz})^2 + (\varepsilon_{zz}-\varepsilon_{xx})^2 + 6(\varepsilon_{xy}^2 + \varepsilon_{yz}^2 + \varepsilon_{zx}^2)\right]^{1/2}, \qquad (2)$$

and the shear stress is calculated according to

$$\sigma = \frac{1}{\sqrt{2}}\left[(\sigma_{xx}-\sigma_{yy})^2 + (\sigma_{yy}-\sigma_{zz})^2 + (\sigma_{zz}-\sigma_{xx})^2 + 6(\sigma_{xy}^2 + \sigma_{yz}^2 + \sigma_{zx}^2)\right]^{1/2}. \qquad (3)$$

These formulas are commonly used in isotropic plasticity [61], derived from the second invariant of the deviatoric strain and stress, respectively. The initially cubic box is compressed along the *x* and *y* directions and expanded in the *z* direction such that the volume is constant. The resulting shear stain is $\varepsilon_{xx}(t)=\varepsilon_{yy}(t)=(-1/2)\varepsilon_{zz}(t) + O(\varepsilon_{zz}^2)$. Using Eq. (2) the strain report is then $\varepsilon(t)=\varepsilon_{zz}(t) + O(\varepsilon_{zz}^2)$. By the end of the simulation the strain in some cases is greater than 1 (>100%), so the higher order corrections are important. The stress, too, is biaxial, specified by $\sigma_{xx}(t)=\sigma_{yy}(t)$ and $\sigma_{zz}(t)$. The pressure in all of the simulations is roughly constant at 2-3 GPa, but not perfectly constant due to the small excess volume in the growing population of lattice defects and the coupling between the hydrostatic and shear deformations in nonlinear elasticity.

We can understand this behavior in more detail, beginning with the elastic phase. Up to a shear strain of 0.02 the deformation is elastic, so the stress and the strain are related by the elastic constants of the polycrystalline system: the bulk modulus *B* and the shear modulus *G*. The microstructure is essentially isotropic and equiaxed, so several continuum-level homogenization techniques are available that relate the polycrystalline shear modulus *G* to the single-crystal elastic constants $C_{ijkl}$ (more precisely, the stress-strain coefficients $B_{ijkl}$ in the case of non-zero pressure [62]). The homogenization techniques solve the equations of mechanical equilibrium and compatibility for an elastic polycrystal based on various simplifying assumptions. Since the elastic anisotropy of tantalum is weak ($A_{Zener}$=1.56) [63], even relatively simple homogenization techniques work well, including the Voigt (constant strain) [64] and Reuss (constant stress) [65] averages, $G_V$ and $G_R$, respectively. More accurate estimates of the polycrystalline shear modulus include the Voigt-Reuss-Hill average shear modulus, $G_{VRH}$, [66], equal to the arithmetic mean of the Voigt and Reuss shear moduli, and the more sophisticated formula due to Hershey, $G_H$ [67]. The single-crystal elastic constants for the Finnis-Sinclair potential we have used are: $B$=196 GPa, $C'=(C_{11}-C_{12})/2$=52.4 GPa, $C_{44}$= 82.4 GPa at $T$=0K [39]. The corresponding polycrystalline shear moduli are $G_V$=70.4 GPa, $G_R$=67.0 GPa, $G_{VRH}$=68.7 GPa and $G_H$=68.9 GPa, also at absolute zero temperature. The difference between the Voigt-Reuss-Hill and Hershey values is only 0.3% due to the low anisotropy, suggesting these estimates should be quite good. The value of the shear modulus calculated from the simulation is $G = \dfrac{1}{3}\dfrac{d\sigma}{d\varepsilon}\bigg|_{\varepsilon=0} = 60.7$ GPa. This value is 12% lower than $G_H(T=0K)$. Some of the difference may be explained by thermal softening, but this decrease is about three times larger than the experimental value for the softening of conventional polycrystalline Ta. A small contribution may also be expected to come from differences in the elastic constants at the grain boundaries and statistical deviations from a perfectly random texture in the microstructure, but we have not attempted to quantify these contributions. In Fig. 2 we show a plot of an elastic curve that we calculated from two single crystal MD stress-strain curves for tetragonal and trigonal shear at $T$=300K, combined through a Voigt average.

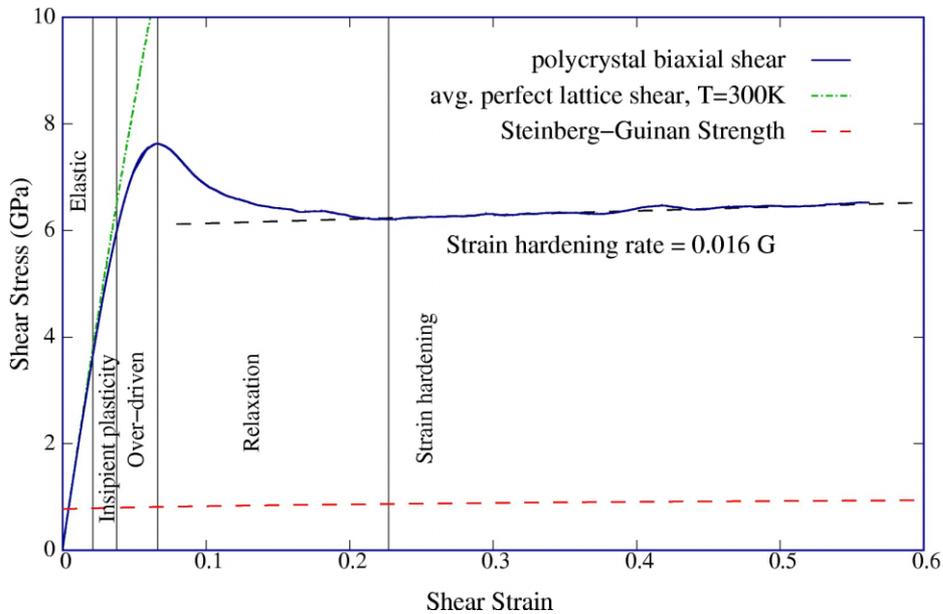

Figure 2. Stress – strain relationship for nanocrystalline Ta as simulated in MD for biaxial compression at a strain rate of $10^9$/s. The other curve is the room temperature single-crystal stress-strain averaged over orientation. The different regimes of the deformation are indicated. For comparison, the high-strain-rate Steinberg-Guinan stress-strain curve at T=300K, P=2GPa is also plotted.

As the shear stress increases, dislocations begin to move, ending the elastic regime. In conventional plasticity theory, there is a yield point followed by plastic deformation until the ultimate strain. More sophisticated models of the plastic behavior, such as the Mechanical Threshold Stress (MTS) model, account for thermally activated dislocation motion as well as the strength of the forest dislocations, and plasticity may begin before what is identified as the macroscopic yield point [68]. In our simulations, the stress-strain curve departs from the elastic stress-strain curve well before the peak flow stress. For example, in Fig. 2 the departure from elasticity is at 3.8 GPa, whereas the peak flow stress is 7.6 GPa. This departure corresponds to the point that a small amount of dislocation motion is evident in visualization of the orientation field. We describe this analysis in more detail below. Throughout this range the plastic deformation rate is increasing due to the rising shear stress, but it is not sufficiently large to compensate for the applied shear strain rate, so the shear stress continues to rise. Orowan's Law provides the kinematic relationship between the dislocation flow as expressed by a density of mobile dislocations, $\rho$, and their mean velocity, $v$, and the dislocation-flow component of the plastic strain rate $\dot{\varepsilon}_p$:

$$\dot{\varepsilon}_p = \rho b v \qquad (4)$$

(see for example Ref. [63]) where $b$ is the Burgers vector ($b = \sqrt{3}/2\,a =$ 2.87Å at ambient conditions, where $a$ is the bcc Ta lattice constant). In this regime of incipient plasticity, the density of mobile dislocations is very low initially, so $\dot{\varepsilon}_p$ is low. As the shear stress increases, more dislocations are emitted from the grain boundaries, so $\rho$ increases and the plastic strain rate increases until it equals the applied strain rate. Dislocation motion is not the only contribution to plasticity. Visualization shows that twinning also occurs in this regime, as shown in Fig. 3. Twinning also contributes to the plastic strain rate and the total plastic strain, limited in magnitude by the crystallography of twinning [63]. The {112}<111> twinning in bcc metals corresponds to a maximum twinning strain of 0.707. In our simulations the twinning volume is much less than 100%, so the twinning contribution to the plastic strain is much less than the theoretical maximum, and most of the plastic strain is due to dislocation flow.

For comparison, the high-strain rate Steinberg-Guinan stress-strain curve for Ta at P=2GPa and T=300K [29] is also plotted in Fig. 2. The Steinberg-Guinan model is based on a scaling of the yield point at ambient conditions by the pressure and temperature dependence of the shear modulus. It also includes strain hardening. The model was parameterized by fitting to dynamic experiments on conventional polycrystalline samples nominally at a strain rate of $10^5$/s. The increased flow stress in the biaxial compression simulation (Fig. 2) results from the Hall-Petch strengthening of the nanocrystalline microstructure and the higher strain rates. At the peak stress, the flow stress in the MD simulation is 9.5 times that of the Steinberg-Guinan model. The Steinberg-Guinan model does not have a peak in its stress-strain curve. Other strength models developed for high-rate deformation, such as the Hoge-Mukherjee [69], Steinberg-Lund [29], and PTW [32] models, also do not have stress peaks. The stress peak results from the system being driven at a rate higher than the plasticity can respond, an effect that is not included in these strength models. This effect is included in plastic flow models like the Gilman model [70] in which the initial dislocation density may not be sufficient to generate a plastic strain rate equal to the total rate of shear strain.

Figure 3 shows the lattice orientation for atoms in a slice of the simulation. The same Lagrangian slice is shown at different strains in the three panels. The grain structure is clearly evident using the templated orientation technique and quaternion color map described above (displayed here in black and white). The atoms are plotted at their locations in real space, and the strain of the simulation box is evident. The microstructure is evolving, with grains subdividing due to twinning and grains coalescing due to rotation by plastic flow. An increase in the aspect ratio of grains is also evident, as the shear flow distorts the initially equiaxed grains. Even though twinning is evident and an important grain refinement process [71], it is limited by the lattice relations described above, and it only contributes a small part of the ultimate shear strain that is attained in the simulation.

The mechanisms of plastic deformation in nanocrystalline materials have been discussed extensively in the literature [5,9,72], including the use of MD simulations to identify mechanisms of grain rotation [3] and dislocation processes associated with the grain boundaries [7,73]. The majority of that work has focused on face-centered cubic (fcc) materials in which partial dislocations play a very important role. A recent paper studied diffusional creep in nanocrystalline bcc molybdenum [74]. In bcc metals including the tantalum studied here, dislocations are not dissociated into partials. In fcc metals the similarity in bond lengths across the near neighbor shells for the fcc and hexagonal close-packed (hcp) structures leads to a small energy splitting between fcc and hcp. For that reason, the intrinsic stacking fault energy is low compared to the elastic repulsion energy between partial dislocations, and the partial dissociated to separations that are not negligible compared to the length scale of nanocrystalline grains. In the bcc structure, there is no comparable relation to the structure of the intrinsic stacking fault, so the fault energies are relatively high and the dislocations, albeit exhibiting polarized core structures, do not dissociate [63].

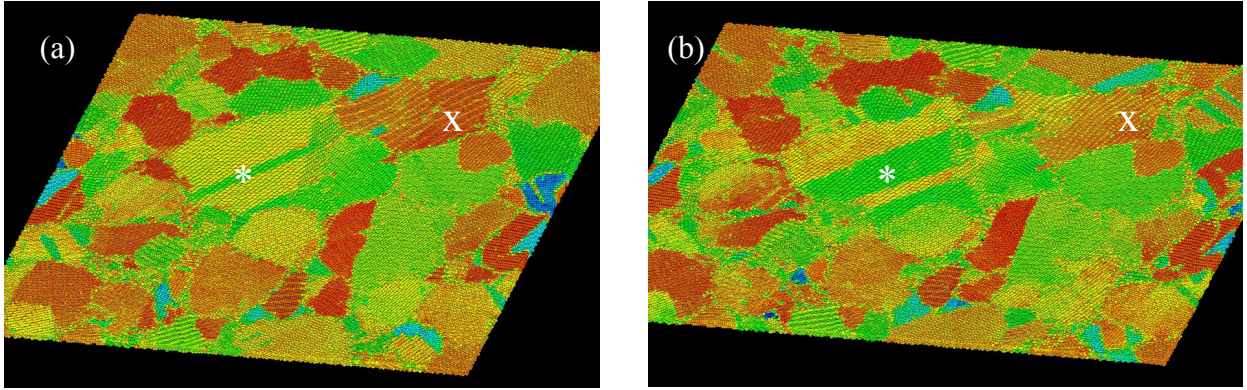

Figure 3. Evolution of the microstructure during biaxial compression at a strain rate of $10^9$/s, as simulated in MD. The two panels show the orientations for each atom in a slice of the simulation, computed and shaded as described in the text. The panels are at strains of (a) 0.13 and (b) 0.45. It is the same slice in both cases (and the same as in Fig. 1), apart from the overall strain. The region marked with an asterisk (*) was a single grain at the start of the simulation and is now undergoing twinning. The mottled pattern of the region marked with an x is due to dislocations, and substantial dislocation flow has led to a rotation of the lattice.

Once the maximum flow stress is attained, the flow stress drops as the deformation continues. This drop in flow stress is approximately exponential, but it is not completely smooth. The plot in Fig. 2 at a strain rate of $1 \times 10^9$/s is fairly smooth, but particularly at lower strain rates there is a marked jagged character to the stress. This punctuation of the stress relaxation with small plateaus and precipitous drops is reminiscent of the serrated flow seen in nanoindentation associated with dislocation cascades [75]. Here, too, it is a manifestation of the relative smallness of the plastic zone, and it would become much less pronounced in a larger simulation. There is no mechanism of plasticity we observe that spans the entire simulation. The nucleation of dislocations and twins is expected to be governed by the overall shear stress modified by local stress concentrators, as well as the usual considerations of lattice orientations (Schmid factors), mobilities and nucleation thresholds [63]. In a larger simulation the statistics of dislocation and twinning events would average out to make the curve more smooth; still the features on the curve are interesting and suggest that grains go through quiescent and active phases as the plastic deformation progresses. We have observed that the plasticity is inhomogeneous in space and time, but a careful analysis of this process remains for future work.

Finally, the stress relaxation phase ends, and the flow stress begins to increase again weakly. The transition occurs at a strain of 0.23 and a stress of 6.2 GPa in Fig. 2. We calculate the strain hardening rate to be $\frac{d\sigma}{d\varepsilon} = 0.016\,G$, expressed in terms of the shear modulus. For comparison Gray and Rollett [76] report a value of G/680 (=0.0015G) for conventional polycrystalline tantalum at a strain rate of $3 \times 10^3$/s; our observed strain hardening rate is an order of magnitude larger.

In the visualization of the simulation, non-dissociated dislocations are observed to nucleate from grain boundaries and rapidly traverse the grain to be absorbed in the grain boundary on the opposite side. This dislocation flow acts to relieve the shear stress as the plastic strain described by Orowan's equation (4) reduces the part of the total shear strain that must be accommodated elastically. In bulk plasticity, a mobile dislocation's glide can carry it into the forest dislocations where it forms junctions and becomes sessile, no longer moving [63]. In the nanoscale grains simulated here, the dislocation density would have to be enormous to have appreciable forest interactions, and it is not. The dislocations undergo ballistic glide from the time they are emitted until they are absorbed. As a result, there is little strain hardening by the conventional mechanism of forest interactions. Instead, the grain boundaries govern the flow, and it is the evolution of the grain microstructure that leads to hardening. The grain boundary network evolves due to the creation of new boundaries through twinning and the reduction in the efficacy of grain boundary impediments through grain rotation. We have observed dislocations being impeded by their interaction with twin boundaries that did not exist at the start of the simulation.

**Uniaxial compression**. We have also investigated the behavior of the nanocrystalline Ta system under volume-conserving uniaxial compression over a range of strain rates. Uniaxial compression is more common in materials dynamics experiments because planar waves are easier to probe and analyze. The results from volume-conserving uniaxial compression simulations are interesting in their own right, and may be contrasted with the results from the volume-conserving biaxial compression simulations to study tension/compression asymmetry [77]. The stress-strain curves are shown in Fig. 4 for strain rates of $3 \times 10^9$/s, $1 \times 10^9$/s, $3 \times 10^8$/s, and $1 \times 10^8$/s.

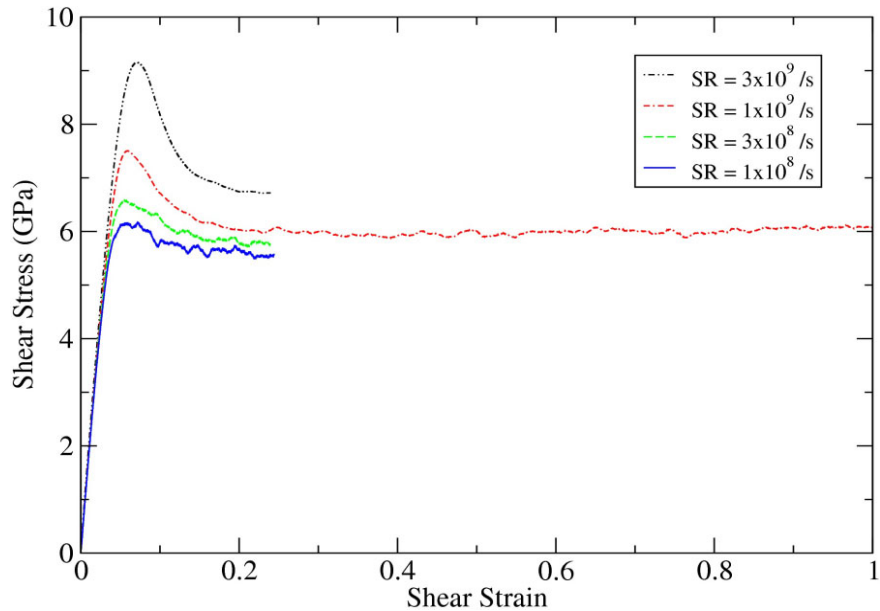

Figure 4. Stress – strain relationship for nanoscrystalline Ta as simulated in MD for uniaxial compression. The four curves correspond to different strain rates (SR), as shown in the legend.

The mechanisms of plasticity in uniaxial deformation are the same as in the biaxial case: dislocation flow and twinning. We have carried out the simulations at several strain rates in order to investigate strain rate effects. The peak stress and the stress at the high-strain plateau are both observed to increase with strain rate. Their values are reported in Table 1.

| Strain Rate | Peak Stress | Plateau Stress |
|---|---|---|
| $3 \times 10^9$/s | 9.2 GPa at $\varepsilon$=0.072 | 6.7 GPa* |
| $1 \times 10^9$/s | 7.5 GPa at $\varepsilon$=0.059 | 5.9 GPa |
| $3 \times 10^8$/s | 6.6 GPa at $\varepsilon$=0.055 | 5.7 GPa* |
| $1 \times 10^8$/s | 6.2 GPa at $\varepsilon$=0.072 | 5.5 GPa* |

Table 1. Peak stress and plateau stress at four different strain rates for the uniaxial deformation of nanocrystalline Ta as simulated in MD. The strain at which the peak stess occurred is also reported. The $\dot{\varepsilon} = 10^8$/s simulation was terminated before it reached the stress plateau, so no platuea value is reported. The asterisks denote the final value from the simulation rather than a minimum.

The peak in the stress occurs at a slightly lower stress than in the biaxial $\dot{\varepsilon}=10^9$ simulation discussed above: 7.5 GPa compared to 7.6 GPa. The early parts of the stress-strain curves are very similar; however, there is compression/tension asymmetry in the strain hardening, as we see below. The peak stress increases with strain rate. We have fit this increase to a form motivated by the Gilman model [70]: $\dot{\varepsilon} = \dot{\varepsilon}^* \exp[A/(\tau_0 - \tau_{peak})]$. Equivalently,

$$\tau_{peak} = \tau_0 + A/\log(\dot{\varepsilon}^*/\dot{\varepsilon}). \tag{5}$$

Here $\tau_0$ is a hypothesized threshold shear stress at which the stress peak first forms and $\dot{\varepsilon}^*$ is a maximum plastic strain rate. The peak stresses shown in Table 1 are fit with the following parameters: $\dot{\varepsilon}^* = 1.69 \times 10^{10}/s$, $\tau_0 = 4.62 GPa$ and $A = 8.02 GPa$. Based on four data points, we cannot make a definitive statement that Eq. (5) is the correct form for strain rates outside the range we have simulated. It is a reasonable form and works well for these simulations.

The trend is for the peak to occur at larger strains for larger strain rates; however, the $\dot{\varepsilon}=10^8$ case does not follow this trend, apparently because of a stress fluctuation at ε=0.072. The strain at the peak stress is roughly fit by $\varepsilon_{peak} = \varepsilon_0 + B(\dot{\varepsilon}/\dot{\varepsilon}_0)\log(\dot{\varepsilon}/\dot{\varepsilon}_0)$ where $\varepsilon_0 = 0.53$, $\dot{\varepsilon}_0 = 10^8/s$ and $B = 0.00063$. The plateau stress, by which we mean the minimum stress following the peak, is also reported in Table 1. We refer to it as a plateau because the strain hardening rate is so small. Only the $\dot{\varepsilon}=10^9$ simulation has been carried out sufficiently far in strain to calculate the strain hardening rate and to determine the plateau stress with certainty; for the other strain rates, we quote the final value of the stress as the plateau, which should be a good approximation due to the low strain hardening. For the $\dot{\varepsilon}=10^9$ simulation we calculate the strain hardening rate to be $\frac{d\sigma}{d\varepsilon} = 0.036 GPa = 0.0005 G$, which is a very small value. Small strain hardening rates are indicative of a propensity for shear localization [78]. This strain hardening rate is a factor of 3 lower than the experimental value quoted above, and a factor of 30 lower than the biaxial value. At this point we have not identified mechanisms that account for the difference.

**Amorphous tantalum and dynamic recrystallization**. In the biaxial and uniaxial compression simulations described above, the initial polycrystalline microstructure is the same: the microstructure that resulted from the pressurization-induced solidification simulations of Streitz et al. [45]. That initial microstructure is quite fine-grained. It is interesting to contemplate how the behavior of the system differs in taking the grain size to the limit of refinement: a system in which there are no crystalline grains: an amorphous system. There is no evidence for a metastable amorphous tantalum at any pressure, but if the crystallization kinetics are sufficiently slow, a supercooled liquid state of tantalum might exist for part or all of the time of interest in a dynamic experiment. Since the solidus line can also be crossed by compression, there might be a supercompressed liquid state as well that behaves like an amorphous solid. Indeed, it is easier to design an experiment for rapid compression than for rapid cooling, so a supercompressed liquid experiment may well be possible.

We investigate the rapid deformation of an amorphous Ta system using the same size system as in the nanocrystalline simulations: 16M atoms in an initially cubic box 66.8 nm on a side with periodic boundary conditions. The initial configuration is attained by heating the system to 10000K and then rapidly quenching it over a period of 350 femtoseconds to room temperature ($T$=300K). We have used the orientation analysis to confirm that the system contains no crystallites (regions where the atomic orientation is correlated beyond nearest neighbors). The system was then held at $T$=300K for 3 ps prior to beginning the deformation.

The stress-strain curves for simulations with an amorphous (quenched liquid) initial configuration are shown in Fig. 5. The curves appear similar in character to the nanocrystalline stress-strain curves. Again, there is an initial elastic regime, a peak in the flow stress and an approximate plateau at large strain. The maximum flow stress and plateau stresses increase with increasing strain rate. It is interesting, however, that the peak stress is substantially lower than in the nanocrystalline case: 4.89 GPa for the amorphous system vs. 7.5 GPa for the nanocrystalline system at a strain rate of $10^9$/s. This reduction in the peak stress is consistent with the inverse Hall-Petch trend of reduction in yield strength with reduction in grain size: the amorphous system is the ultimate limit in which the entire system is filled with material similar to a random grain boundary [2]. Also, it is noteworthy that the system undergoes strain softening: as the strain increases past the point of peak stress, the flow stress decreases. Unlike the nanocrystalline system, the flow stress never reaches a minimum out to strains of 100%. The strain hardening rate is -0.23 GPa = -G/240, where the negative sign indicates softening and G~55GPa for the amorphous system. Strain softening is often indicative of a shear localization instability, since regions that deform more become softer and are then more easily deformed even further [78]. Since the system does harden with increasing strain rate, the degree of localization in dynamic deformation depends on whether a localized region with increased strain and strain rate is softer than the surroundings.

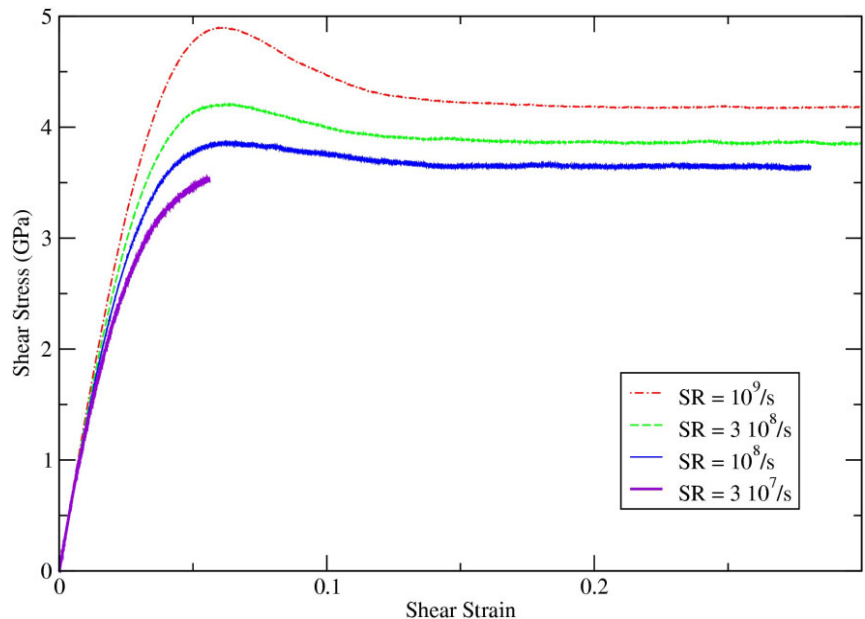

Figure 5. Stress – strain relationship for amorphous (quenched liquid) Ta as simulated in MD for uniaxial compression. The four curves correspond to different strain rates (SR), as indicated in the legend.

In the case of the nanocrystalline system we observed changes in the microstructure that were associated with the plastic flow. In the amorphous system, the changes are more subtle. For much of the initial phase of the deformation, no change to the amorphous structure is evident; however, late in time, crystallite nucleation is observed to begin. Consider the two snapshots shown in Fig. 6. In the first snapshot (Fig. 6a) at the beginning of the simulation no crystallites are observed. In the second snapshot (Fig. 6b) at a strain of 0.87 in the $\dot{\varepsilon}=10^9$/s simulation, several crystallites are evident. At this point ~1% of the simulation has crystallized. We do not yet have a model for how the crystallite formation affects the flow stress. Strain softening occurs prior to substantial recrystallization, so it is not clear how the two are related, if they are. This process is clearly different than the recrystallization of tantalum that has been studied previously [79].

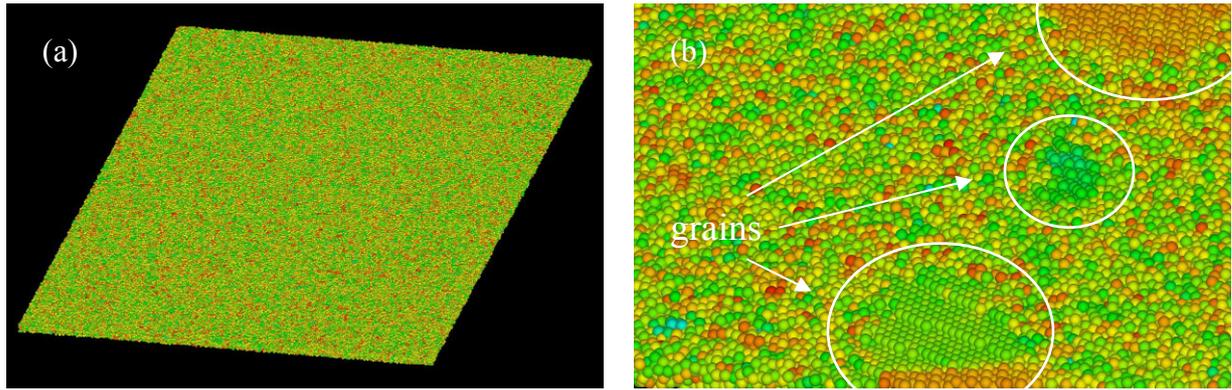

Figure 6. Snapshots of a slice of the (a) initial and (b) t=0.87 ns microstructures of the initially amorphous (quenched liquid) Ta as simulated in MD. The formation of grains from the amorphous matrix for recrystallization is evident in the circled regions in panel (b). The view in panel (b) is magnitifed 5x with respect to panel (a).

**Discussion and Conclusion**

We have used MD to examine the plastic flow behavior in nanocrystalline and amorphous (quenched liquid) systems of Ta. In all of the systems and at all of the strain rates studied, a general stress-strain behavior is evident. An initial elastic regime is followed by a peak in the flow stress due to plasticity and then a reduction in the flow stress as the deformation continues. In the nanocrystalline system, the flow stress reaches a minimum and then increases with further deformation; in the amorphous systems, no subsequent minimum was attained—the flow stress continued to decrease. In the nanocrystalline system, the uniaxial and biaxial simulations exhibited roughly the same peak stress, but the strain hardening rate was significantly higher in the biaxial compression case. This effect is reminiscent of compression/tension asymmetries studied in other types of plastic deformation of bcc metals, but the exact mechanism has yet to be determined.

The amorphous system is, at least in an approximate sense, a nanocrystalline system in the limit of zero grain size. The entire system is similar to the material in random grain boundaries. Here "random" is used in the sense of not special (not a coincident site lattice boundary). It would be interesting to investigate a broader range of nanocrystalline systems, and quantify the trends in peak stress and strain hardening rate as a function of grain size.

We used a local lattice orientation characterization to investigate changes in the microstructure due to the deformation. In the nanocrystalline systems, both dislocation flow and twinning were observed. The twinning reduced the size of large grains, whereas in some cases significant amounts of dislocation flow led to grain rotation and coalescence. In the case of the quenched liquid system, the nucleation of grains from the amorphous matrix was observed in the late stages of deformation. The lattice orientation characterization has provided a great deal of insight about the plastic deformation mechanisms in the larger grains. It is less well suited to study the deformation of the smaller grains in the size distribution, and there are many of them. The characterization of the plasticity in these small grains is an important problem for future work.

These simulations lead to many other questions. Ultimately, comparison with experiment is essential, and laser and Z-pinch facilities are making great strides at producing high rate deformation of metals without shock heating. The experiments are challenging, especially with regard to *in-situ* characterization of the material response, and it may be some time before large laser facilities can attain detailed information about the mechanisms of high rate deformation. In the mean time, much more can be done with simulation. A more detailed investigation of the nanocrystalline and quenched liquid systems could answer many remaining questions. We have not investigated the stability of these systems in the absence of deformation, and this control simulation needs to be done, especially for the quenched liquid. How are the time scales for recrystallization in the presence and absence of deformation related? A model to account for the strain softening would be very helpful. Even more interesting would be a predictive model of the phases of flow for the nanocrystalline system, one that accounts for the interaction of the plastic flow and twinning.

Ultimately, as computer resources increase, it would be very interesting to do an explicit ramped wave simulation of the plasticity of nanocrystalline Ta, along the lines of Ref. [38] but with lower, more realistic strain rates. Because of the combined needs of large systems and long simulated times, as mentioned above, this goal is beyond our current capabilities, but perhaps an advanced technique, such as concurrent multiscale modeling [43,80], can be used to accelerate the simulation. Eventually, it may even be possible to do a simulation in which the Ta begins molten, and as a ramp wave passes it solidifies into a nanocrystalline system and undergoes plastic flow in the continuing compression. Since the wave structure distinguishes the longitudinal direction from the transverse directions, the rotational symmetry of the system is broken and a textured microstructure may result. The shear flow may also lead to crystallites that acquire a non-trivial aspect ratio during the solidification. At this point, these features are just speculation, but as computer power increases it may be possible to conduct the simulations; indeed, as experimental capabilities increase, it may be possible to use techniques such as Small-Angle X-ray Scattering (SAXS) [81] to observe the process *in situ*. There is much to be learned still.

## Acknowledgments

It is a pleasure to thank Bruce Remington, Fred Streitz, Christine Wu, Kyle Caspersen, Meijie Tang, Rich Becker, Bryan Reed, Mukul Kumar, Dave Richards, Rip Collins, Ray Smith and Jon Eggert for useful discussions. The initial atomic configuration for the nanocrystalline simulations was provided by Streitz, Glosli, and Patel [45]. This work was funded in part by the LLNL Laboratory Directed Research and Development (LDRD) program through grant 09-SI-010. Supercomputer resources were provided by Livermore Computing. This work was performed under the auspices of the US Department of Energy by Lawrence Livermore National Laboratory under Contract DE-AC52-07NA27344.